\documentstyle[12pt]{article}
\newcommand{\EQ}{\begin{equation}}
\newcommand{\EN}{\end{equation}}
\newcommand{\bea}{\begin{eqnarray}}
\newcommand{\eea}{\end{eqnarray}}
\newcommand{\hs}{\hspace{0.1cm}}

\newcommand{\th}{\theta}
\newcommand{\lab}{\label}

\newcommand{\var}{\varepsilon}
\newcommand{\p}{\partial}
\newcommand{\al}{\alpha}
\newcommand{\goto}{\rightarrow}
\newcommand{\be}{\beta}

\begin{document}
\setcounter{page}{0}
\topmargin 0pt
\oddsidemargin 5mm
\renewcommand{\thefootnote}{\arabic{footnote}}
\newpage
\setcounter{page}{0}
\begin{titlepage}
\begin{flushright}
ISAS/EP/94/152
\end{flushright}
\vspace{0.5cm}
\begin{center}
{\large {\bf Correlation Functions Along a Massless Flow}} \\
\vspace{1.8cm}
{\large G. Delfino, G. Mussardo and P. Simonetti}\\
\vspace{0.5cm}
{\em International School for Advanced Studies,\\
and \\
Istituto Nazionale di Fisica Nucleare\\
34014 Trieste, Italy}\\

\end{center}
\vspace{1.2cm}

\renewcommand{\thefootnote}{\arabic{footnote}}
\setcounter{footnote}{0}

\begin{abstract}
\noindent
A non-perturbative method based on the Form Factor bootstrap approach is
proposed for the analysis of correlation functions of 2-D massless integrable
theories and applied to the massless flow between the Tricritical and the
Critical Ising Models.
\end{abstract}

\vspace{.3cm}

\end{titlepage}

\newpage

\noindent
1. The 2-D integrable Quantum Field Theories (QFT) associated to the massless
Renormalization Group (RG) flows between two different fixed points describe
important physical situations and present several reasons of interest.
Relevant examples are provided by the $O(3)$ non-linear sigma model with the
$\theta=\pi$ topological term \cite{sigma}, by the massless flows between two
consecutive conformal minimal models \cite{min} and by the 2-D formulation of
the Kondo problem \cite{K}. One of the important features of these QFT
(in addition to their integrability) is the absence of a mass gap, i.e.
their correlation functions have a power law behaviour both in the
ultraviolet and in the infrared regions (with different critical
exponents, though), separated by a non trivial crossover in between.
Usual perturbative methods may fail to capture the physical content of this
kind of theories and it is therefore useful to develop some alternative
approach to their analysis. Such an approach does exist for the massive QFT,
where the powerful method of the Form Factors (FF) \cite{KWS} allows
us to enlighten  their dynamics and provides, in particular, fast convergent
series for the correlation functions \cite{KWS,CM,FF}. In the light of these
achievements, it is an important question to see whether or not
the FF approach can be also successfully applied to the massless
QFT. There are, however, several aspects that need a
careful investigation, in particular the analytic structure of the FF and the
convergent properties of the spectral representation series. The purpose of
this letter is to study the simplest massless non-scaling invariant theory
associated to the flow from the Tricritical Ising Model (TIM) to the
Critical Ising Model. A suitable extension of this approach may be useful to
investigate a concrete physical problem, as for instance the computation
of correlators of the Kondo problem along the flow between the two fixed
points.

\vspace{3mm}

\noindent
2. The massless scattering theory has been developed in \cite{sigma,ZTIM,FSZ}.
In a massless 2-D theory, the excitations consist in a set of left
(L) and right (R) movers. These are the lowest energy states which
propagate all along the RG flow and constitute the relevant degrees of
freedom of the problem. Higher massive states are regarded as decoupled,
their influence being eventually seen in the analytic structure of
the physical amplitudes of the massless particles. The simplest way
of approaching the massless integrable scattering theory is probably to think
of it as a particular limit of a factorizable massive scattering theory.
To avoid inessential complications, let us consider a massive integrable
model whose spectrum consists of a single self-conjugate particle $A$ of mass
$m$ which, in a suitable limit, gives rise to a massless theory with
right-mover and left-mover particles. A technical important point is the
different role played by the rapidity variable in the massive and in the
massless theories. In the massive theory, the dispersion relation may be
written in terms of the rapidity as
\EQ
p^0 = m\cosh\th \,\,\,\,\,, \hspace{5mm} p^1 = m\sinh\th \,\,\, .
\lab{rap}
\EN
The elastic two-particle $S$-matrix in the massive theories only depends on
the Mandelstam variable $s$, given by
$s=\left(p^\mu_1(\th_1)+p^\mu_2(\th_2)\right)^2 = 2 m^2(1 + \cosh\th)$,
where $\th \equiv \th_1-\th_2$. Use of the rapidity variable substantially
simplifies the analytic structure of the amplitude by absorbing the
threshold branch cut singularities in the original variable $s$.
The physical sheet of the $s$-plane can be mapped into the strip
$0\leq$ Im $\th\leq\pi$ of the $\th$-plane in such a way that the two-particle
$S$-matrix becomes a meromorphic function of the rapidity difference.
In a massless theory, there are two possible scattering configurations. The
first of them consists in the RL scattering, the second in the RR (LL)
scattering. In the RL case, the Mandelstam variable $s$ is the only
independent relativistic invariant quantity. Since the
mass gap is zero, the unitarity and crossing cuts of the $S$-matrix in the
$s$-plane join at the origin, so that the Riemann surface splits into two
distinct parts: the ``upper'' (``lower'') part consists in the half of the
physical (unphysical) sheet with $Im\,s>0$ and in the half of the unphysical
(physical) sheet with $Im\,s<0$.
Some care is required to introduce a parameterization in
terms of rapidities since we need to send $m$ to zero in eq.\,(\ref{rap}) but
keeping the energy and the momentum finite. For this aim, it is sufficient to
replace $\th$ by $\be\pm \be_0/2$ and to take the limits
$m\goto 0,\be_0\goto +\infty$ in such a way that the mass parameter
$M\equiv me^{\be_0/2}$ remains finite. According to the sign in front of
$\be_0$, we have
\bea
&& p^0 \,=\,\,\,p^1 \,=\,\,\,\frac{M}{2}e^\be\hspace{1cm}\mbox{\,\,\,for
right-movers}\,\,,
\lab{mrap}\\
&& p^0\,=\,-p^1\,=\,\frac{M}{2}e^{-\be}\hspace{1cm}\mbox{for
left-movers}\,\,\,.
\nonumber
\eea
In short, it is useful to regard a right-mover $A_R(\be)$ or a
left-mover $A_L(\be)$ as formally defined by the limits
$A_{R,L}(\be)=\lim_{\be_0\goto +\infty}A(\be \pm\be_0/2)$.
The Mandelstam variable $s$ is now given by $ s\,=\,M^2e^{\be_1-\be_2}$.
Due to the aforementioned splitting of the $s$-plane,
in the massless case we need two meromorphic functions of
the rapidity difference in order to represent a function defined on the two
sheets of the Riemann surface.
Let $S_{RL}(\be)$ ($\tilde{S}_{RL} (\be)$) be
the values of the scattering amplitude on the ``upper'' (``lower'') sheet
above defined. $S_{RL}(\be)$ ($\tilde{S}_{RL}(\be)$) is
obtained from the value $S(\th)$ ($S(-\th)$) of the massive amplitude
as the limit $A(\th_1)\goto A_R(\be_1)$ and $A(\th_2)\goto A_L(\be_2)$, i.e.
\[
S_{RL}(\be) = \lim_{\be_0\goto +\infty} S(\be + \be_0)
\,\,\, ,
\hspace{3mm}
\tilde{S}_{RL}(\be) = \lim_{\be_0\goto +\infty} S(-\be-\be_0)\,\,\,.
\]
The unitarity and crossing relations can be now expressed as
$ S_{RL}(\be) \,\tilde{S}_{RL}(\be)\,=\,1$ and
$S_{RL}(\be)\,=\,\tilde{S}_{RL}(\be+i\pi)$.
Let us now turn our attention to the
RR (LL) scattering. Since in these channels the variable $s$ is identically
zero, all the usual analyticity arguments of the $S$-matrix theory cannot
apply, a circumstance which clearly reflects the lacking of intuitive
understanding of the massless scattering process in ($1+1$) dimensions.
In spite of this fact, the amplitudes $S_{RR}$ and $S_{LL}$ can be formally
defined in terms of the massless limit of the massive theory. However, the
rapidity shifts in the massless limit cancel in these cases and
therefore we simply have $ S_{RR}(\be)=S_{LL}(\be)=S(\be)$. In view of this
identity, the amplitudes $S_{RR}$ and $S_{LL}$ are expected to fulfill (in the
rapidity variable) the equations valid for the massive theory. As in the
massive case, it is useful to encode all the scattering information of the
massless theories into the Faddev-Zamolodchikov algebra
\[
A_{\alpha_1}(\be_1)A_{\alpha_2}(\be_2)=S_{\alpha_1\alpha_2}(\be_1-\be_2)
A_{\alpha_2}(\be_2)A_{\alpha_1}(\be_1)\,\,,
\lab{mfz}
\]
where $\alpha_i=R,L$, $A_R(\be)$ and $A_{L}(\be)$ are the creation operators
for the R and L movers respectively and
$S_{LR}(\be_1-\be_2)=\tilde{S}_{RL}(\be_2-\be_1)$.

\vspace{3mm}

\noindent
3. Massless form factors are defined to be the matrix elements of local
operators between asymptotic states. To determine their analytic
properties, let us firstly consider the matrix element of a scalar
operator $O(x)$ between the vacuum and the two-particle state
$\mid A_R(\beta_1) A_L(\beta_2)>$. Let $F_{RL}(\be)$ and
$\tilde{F}_{RL}(\be)$ be the functions which, for real
values of $\beta=\beta_1-\beta_2$, take values on the upper and lower edges of
the unitarity cut, respectively. They are related each to the other through
the equations
\bea
&& F_{RL}(\beta)\,=\, S_{RL}(\beta)\tilde{F}_{RL}(\beta)
\,\,\,;\\ \nonumber
&& F_{RL}(\be +i\pi) \,=\, \tilde{F}_{RL}(\be - i\pi)\,\, .
\eea
Using now $F_{LR}(\be)=\tilde{F}_{RL}(-\be)$, and the fact
that $F_{RR}(\be)$ and $F_{LL}(\be)$
share the same properties of the FF of the massive theory, the
monodromy equations for the massless two-particle FF may be written
in a compact form as
\EQ
\begin{array}{l}
F_{\alpha_1\,\alpha_2}(\be)  \,=\, S_{\alpha_1\alpha_2}(\be)
F_{\alpha_2\,\alpha_1}(-\be)\,\,\,;\\
F_{\alpha_1\,\alpha_2}(\be + 2\pi i) \,=\,
F_{\alpha_2\,\alpha_1}(-\be)
\,\, ,
\end{array}
\lab{watwo}
\EN
where $\alpha_i=R,L$. These equations are easily generalized to the
$n$-particle FF,
\bea
&& F_{\al_1\ldots\al_i\al_{i+1}\ldots\al_n}
   (\be_1,\ldots,\be_i,\be_{i+1},\ldots,\be_n) = \nonumber\\
&& \,\,\,\,\,= \,S_{\al_i\al_{i+1}}(\be_i-\be_{i+1})
   F_{\al_1\ldots\al_{i+1}\al_{i}\ldots\al_n}
(\be_1,\ldots,\be_{i+1},\be_{i},\ldots,\be_n)
\label{mw1} \,\,\,;\\
&& F_{\al_1\al_2\ldots\al_n}(\be_1+2\pi i,\be_2,\ldots,\be_n) =
F_{\al_2\ldots\al_{n}\al_1}
(\be_2,\ldots,\be_{n},\be_1)\,\, , \nonumber
\eea
where $F_{\al_1\ldots\al_n}(\be_1,\ldots,\be_n)\,=\,
<0|O(0)|A_{\al_1}(\be_1),\ldots,A_{\al_n}(\be_n)>$ are meromorphic functions
of rapidities defined in the strips $0\leq$ Im $\beta_i<2\pi$.
In the massive case, the form factors present simple pole singularities
associated either to the bound states in the scattering amplitudes or to the
particle-antiparticle annihilation processes \cite{KWS}. However, stable bound
states are usually forbidden in massless theories due to the absence of
thresholds. This leads to exclude the presence of the first kind of
singularities in the RL sub-channels, which are the only ones to which standard
$S$-matrix theory can be applied. On the other hand, the RR and LL
scattering amplitudes formally behave as in the massive case.
Hence, it is natural to assume that, whenever the amplitude $S_{\al\al}(\be)$
has a simple pole for $\be=iu$ ($u\in (0,\pi)$) with residue $g$, this
induces a corresponding equation for the FF
\[
i\,res_{\be_{n+1}-\be_{n+2}=iu}F_{\al_1\ldots\al_n\al\al}(\be_1,\ldots,\be_n,
\be_{n+1},\be_{n+2})=gF_{\al_1\ldots\al_n\al}(\be_1,\ldots,\be_n,
\be_{n+1}-iu)\,\,\,.
\]
With reference to the annihilation poles, they may only occur in the RR or LL
sub-channels. The residue equation in this case reads
\EQ
-i\,res_{\be'=\be+i\pi}F_{\al\al\al_1\ldots\al_n}
(\be',\be,\be_1,\ldots,\be_n) =
 \left(1\,-\,\prod^n_{i=1}S_{\al_i\al}(\be_i-\be)\right)
F_{\al_1\ldots\al_n}(\be_1,\ldots,\be_n)\,\,\,.
\label{resk}
\EN
The invariance of the massless theory under the spatial inversion
transformation will provide the additional relation
$F_{\al_1\ldots\al_n}(\be_1,\ldots,\be_n)=
F_{{\cal P}[\al_n]\ldots {\cal P}[\al_1]}(-\be_n,\ldots,-\be_1)\,\,,
$, where ${\cal P}[R]=L$, ${\cal P}[L]=R$.

\vspace{3mm}

\noindent
4. A perturbation of the TIM by its sub-leading energy operator $\var'$ of
conformal dimensions $(3/5,3/5)$ induces a massless flow that ends to the
Critical Ising Model \cite{min,KMS}. This flow takes place
along the self-dual line of the phase diagram analysed in \cite{ZB,KMS} and
a low energy effective Lagrangian of the corresponding QFT is given by
\EQ
{\cal L}_{\mbox{eff}}\,=\,\psi \bar{\p} \psi \,+\,
\bar{\psi} \p \bar{\psi} \,-\,
\frac{4}{M^2}\,(\psi \p \psi) (\bar{\psi} \bar{\p} \bar{\psi})
\,+\,\ldots
\,\,\,.\lab{effective}
\EN
This Lagrangian describes the critical Ising model (free massless Majorana
fermion) perturbed by the irrelevant operator $(\psi\p\psi)(\bar{\psi}
\bar{\p}\bar{\psi})\sim T\bar{T}$, which is the lowest dimension nonderivative
field invariant under $Z_2$ and duality transformations. Discrete symmetries
of the theory are $\psi\goto\psi,\bar{\psi}\goto-\bar{\psi}$ and
$\psi\goto-\psi,\bar{\psi}\goto-\bar{\psi}$ which can be identified as
the duality transformation and the spin reversal, respectively.
The factorized massless scattering theory of this flow has been proposed in
\cite{ZTIM}. The basic assumption is that the massless neutral fermions
appearing in (\ref{effective}) are the only stable particles of the theory,
i.e. the spectrum is the same as the infrared fixed point theory. Since
$S_{RR}$ and $S_{LL}$ cannot vary along the flow, they are given by the
commutation relations of the Ising fermionic fields,
i.e. $S_{RR}(\be)=S_{LL}(\be)=-1$, whereas for $S_{RL}$ we have
$S_{RL}(\be)=\tanh\left(\frac{\be}{2}-\frac{i\pi}{4}\right)$.
Hence, on the ``upper'' sheet of the $s$-plane we have
$S(s)=\frac{s-iM^2}{s+iM^2}$. Using the unitarity equation,
this expression implies that, while the physical sheet is free of poles, the
unphysical one contains two poles at $s=\pm iM^2$ which can be interpreted as
resonances. Let us turn now to the computation of the
FF\footnote{In the sequel,
$
F_{r,l}(\be_1,..,\be_r;\be'_1,..,\be'_l) =
<0|O(0)|A_R(\be_1)\ldots A_R(\be_r) A_L(\be'_{1})\ldots A_L(\be'_{l})>
$. When $\be_1 > \be_2 >\ldots > \be_r$ and $\be'_1 > \be'_2 >
\ldots > \be'_l$, these functions coincide with the physical matrix
elements computed on the {\em in}-asymptotic states.} of a local operator
${\cal O}$. They can be parameterized as
\bea
&& F_{r,l}(\be_1,\be_2,\ldots,\be_r;\be'_1,\be'_2,\ldots,\be'_l)\,=\,
H_{r,l}\, Q_{r,l}(x_1,x_2,\ldots,x_r;y_1,y_2,\ldots,y_l)\times
\nonumber \\
& & \hspace{15 mm} \times \,
\prod_{1\leq i<j\leq r}\frac{f_{RR}(\be_i-\be_j)}{x_i+x_j}\,
\prod_{i=1}^r \prod_{j=1}^l f_{RL}(\be_i-\be'_j)\,
\prod_{1\leq i<j\leq l}\frac{f_{LL}(\be'_i-\be'_j)}{y_i+y_j}\,\,,
\label{param}
\eea
where $x_i\equiv e^{\be_i}$, $y_i\equiv e^{-\be'_i}$ and $H_{r,l}$ are
normalization constants. The auxiliary functions $f_{RR}$, $f_{LL}$ and
$f_{RL}$ completely take into account the monodromy properties of
the FF (\ref{param}). They are the minimal solutions of the
equations $f_{\al_1\al_2}(\be)\,=\,S_{\al_1\al_2}(\be)
f_{\al_1\al_2}(\be+2\pi i)
$, with neither poles nor zeroes
in the strip $0<$ Im$ \be <2\pi$. Explicitly,
$f_{RR}(\be)\,=\,f_{LL}(\be)\,=\,\sinh\frac{\be}{2}$, and
\[f_{RL}(\be)\,=\,\exp
\left(
\frac{\be}{4}- \int_0^{\infty}\frac{dt}{t}
\frac{\sin^2\left(\frac{(i\pi-\be)t}{2\pi}\right)}{\sinh t \cosh
\frac{t}{2}}
\right)\,\,.
\]
The function $f_{RL}(\be)$ also satisfies the equation
$
f_{RL}(\be\pm i\pi) f_{RL}(\be)\,=\, \frac{i \gamma}{1\pm
ie^{-\be}}\,\,,
$
where $\gamma=\sqrt{2} e^{2 G/\pi}$, $G$ being the Catalan constant.
Since $S_{RR}$ and $S_{LL}$ in this theory
are free of poles, the FF only present kinematical poles
which are explicitly inserted in (\ref{param}) through the factors
$x_i+x_j$ and $y_i+y_j$ in the denominator. With the requirement that the FF
are power bounded in the momentum variables, $Q_{r,l}$ have to be rational
functions, separately symmetric in the $\{x_i\}$ and $\{y_i\}$ with at most
poles located at $x_i=0$ or $y_i=0$.
Inserting the parameterization (\ref{param}) into the residue equations
(\ref{resk}) and using
$H_{r,l} = - i^{-r} 2^{-2r-1} \gamma^l H_{r+2,l} = - i^{-l} 2^{-2l-1}
\gamma^r H_{r,l+2}$,
the recursive equations for $Q_{r,l}$ are given by
\bea
Q_{r+2,l}(-x,x,x_1,\ldots,x_r;y_1,\ldots,y_l)=x^{r-l+1}
\frac{\rho_r}{\lambda_l}{\sum^l_{k=0}}'
(-i x)^k \lambda_k Q_{r,l}(x_1,\ldots,x_r;y_1,\ldots,y_l)
\label{pinch} && \\
Q_{r,l+2}(x_1,\ldots,x_r;y_1,\ldots,y_l,y,-y)=y^{l-r+1}
\frac{\lambda_l}{\rho_r} {\sum^r_{k=0}}'
(-i y)^k \rho_k Q_{r,l}(x_1,\ldots,x_r;y_1,\ldots,y_l) \,\, ,&&\nonumber
\eea
where the primed sums run over odd indices if $(r+l)$ is even and vice versa,
and $\rho_k$ ($\lambda_l$) are the elementary symmetric
polynomials in the variables $\{x_i\}$ ($\{y_i\}$).

\vspace{3mm}

\noindent
5. Eqs.\,(\ref{pinch}) are quite general since they were obtained without
reference to any particular operator. Here we will restrict our attention
to the FF of some operators of particular physical relevance, namely the
trace of the stress-energy tensor $\Theta(x)=T^{\mu}_{\mu}(x)$, the order
and the disorder operators $\Phi(x)$ and $\tilde\Phi(x)$
of conformal dimensions $(3/80,3/80)$ in the ultraviolet limit and
$(1/16,1/16)$ in the infrared one. Since these
fields are spinless, under a Lorentz transformation they satisfy
$
Q_{r,l}(\{e^{\Lambda}x_i\};\{e^{-\Lambda}y_i\})=
e^{\left(\frac{r(r-1)}{2}-\frac{l(l-1)}{2} \right)\Lambda}
Q_{r,l}(\{x_i\};\{y_i\})
$.
Selection rules for the FF are obtained by assigning the transformation
properties of the massless particles under the discrete symmetries of the
theory. According to the invariances of the fermionic Lagrangian
(\ref{effective}), we assign odd $Z_2$-parity to both right and
left-movers, and even (odd) parity to right (left) movers under duality
transformation. The trace of the energy-momentum tensor is expressed in terms
of the ultraviolet perturbing field and the conjugate coupling constant
$\lambda$ by the relation
$ \Theta(x)=\frac{8}{5}\pi{\lambda}\var'(x)$.
Since the subenergy $\var'$ is even under both spin reversal and
duality transformation,  $\Theta$ will have nonvanishing form factors
$F_{r,l}$ only for even $r$ and $l$, starting from $r=l=2$ (the vacuum
expectation value $F_{0,0}$ must be identically zero since it vanishes
in the infrared limit). The conservation of the
energy-momentum tensor implies the factorization
$
Q_{r,l}(\{x_i\};\{y_i\})\,=\,
\rho_1 \lambda_1
T_{r,l}(\{x_i\};\{y_i\})
$.
The leading infrared behaviour of $F_{2,2}$ is easily computed by
using the Lagrangian (\ref{effective})
\EQ
F_{2,2}(\be_1,\be_2;\be'_1,\be'_2)\,\goto\,
- 4\pi M^2 \sinh\frac{\be_1-\be_2}{2}\,\sinh\frac{\be'_1-\be'_2}{2}\,
e^{\be_1+\be_2-\be'_1-\be'_2}\,\, ,
\lab{leading}
\EN
and, with this extra piece of information, we can completely fix its exact
expression
\EQ
F_{2,2}(\be_1,\be_2;\be'_1,\be'_2)\,=\,
\frac{4\pi M^2}{\gamma^2} \sinh \frac{\be_1-\be_2}{2}
\prod_{i,j=1,2} f_{RL}(\be_i-\be'_j)
\sinh \frac{\be'_1-\be'_2}{2}\,\,.
\label{f22}
\EN
The recursive equations (\ref{pinch}) are then iteratively solved by
using (\ref{f22}) as initial condition. Using
$
H_{2n,2m}\,=\,
\pi M^2 i^{n(n+1)+m(m+1)}
2^{2(n^2+m^2)-n-m} \gamma^{-2nm}
$, the first right chains are determined to be
\bea
&& T_{2n,2}(\{x_i\};\{y_i\})=(i)^{n^2-1}\,
   \left( \rho_{2n} \frac{\lambda_1}{\lambda_2} \right)^{n-1}\,\,,
   \nonumber \\
&& T_{2n,4}(\{x_i\};\{y_i\})=(i)^{n^2-2n}\,
   \left(\frac{\rho_{2n}}{\lambda_4}\right)^{n-2}\,
   \sum_{k=0}^{n-1} \rho_{2k+1} \lambda_1^{n-1-k} \lambda_3^k \,\,,
   \label{solutionteta}
\\
&& \,\,\,\,\vdots       \nonumber
\eea
Using now the right-left symmetry relation
$T_{r,l}(\{x_i\};\{y_i\})=T_{l,r}(\{y_i\};\{x_i\})$, one can immediately
obtain the solution for the corresponding left chains. Let us now consider
the magnetization operator. The invariance of the theory under spin reversal
implies that $\Phi$ has nonvanishing FF only on an odd number of
particles whereas $\tilde\Phi$ only on an even one. Taking
$F_{1,0}=F_{0,1}=1$ as
initial conditions for the recursive equations for $\Phi$ and
$F_{0,0}=F_{1,1}=1$ for $\tilde\Phi$, we find\footnote{As explained in the
first reference of \cite{CM}, the recursive equations for the FF of
the disorder operator $\tilde\Phi$ are given by (\ref{resk}) but with a plus
sign in front of the product of $S$-matrices.}
\bea
&& Q_{r,0}=\rho_r^{(r-1)/2}\,\,,\nonumber\\
&& Q_{r,1}=\frac{\rho_r^{r/2-1}}{\lambda_1^{r/2}}\,\,,\nonumber\\
&& Q_{r,2}=\rho_r^{(r-3)/2}\sum_{k=0}^r { }'\rho_k\lambda_2^{(k-r+1)/2}\,\,,
\label{ffsigma}
\\
&& Q_{r,3}=\frac{\rho_r^{(r/2-2)}}{\lambda_3^{r/2-1}}\sum_{k=0}^r { }'
   \rho_k\lambda_2^{k/2}\,\,,\nonumber\\
&& \,\,\,\,\vdots \nonumber
\eea
where the prime denotes a sum only over even indices and $r$ should be chosen
such that $r+l$ is odd for $\Phi$ and even for $\tilde\Phi$.

\vspace{3mm}

\noindent
6. The spectral representation of the two-point functions is given by
\bea
&& <O(x) O(0)>=\sum_{r,l=0}^\infty\frac{1}{r!\,l!}\int_{-\infty}^{+\infty}
   \frac{d\be_1\ldots d\be_r d\be'_1\ldots
   d\be'_l}{(2\pi)^{r+l}}|F_{r,l}(\be_1,\ldots,\be_r;\be'_1\ldots\be'_l)|^2
\nonumber\\
&& \,\,\,\,\,\,\,\,\,\,\times\,\exp\left[-\frac{Mr}{2}
   \left(\sum_{j=1}^r e^{\be_j}+\sum_{j=1}^l
   e^{-\be'_j}\right)\right]\,\,,
\lab{mcorr}
\eea
where we used euclidean invariance to set $x=(ir,0)$. This expression
clearly shows that, contrarily to the massive case, the convergence
in the infrared limit $\be_i\goto-\infty,\be'_i\goto+\infty$ is no longer
guaranteed by the exponential factor inside the integrals and completely
relies on the behaviour of the form factors $F_{r,l}$ in this limit.
Let's first analyze the 2-point function of $\Theta$. As explicitly shown by
eq.\,(\ref{leading}), $F_{2,2}$ goes exponentially to zero in the infrared
limit so that the 4-particle contribution of
$G_{\Theta}(r)=<\Theta(r)\Theta(0)>$ is convergent. It is also easy
to check that the terms with higher number of particles are convergent
and sub-leading with respect to the 4-particle contribution. Then, plugging
expression (\ref{leading}) into the integral (\ref{mcorr}), we find for
$r\rightarrow \infty$, $G_{\Theta}(r)\,\simeq\,\frac{16}{\pi^2M^4r^8}$, which
perfectly matches with the expected infrared power law behaviour. On the other
hand, the behaviour of this function in the ultraviolet limit $r\rightarrow 0$
is fixed by the conformal OPE, i.e.
$
G_{\Theta}(r)\,\simeq\,M^4\left(\frac{8}{5}\pi\al\right)^2(Mr)^{-12/5}
$, where $\al=0.148695516$ \cite{ZTIM}. A logarithmic plot of
$G_{\Theta}(r)$ obtained by including the first two contributions
in the spectral representation (\ref{mcorr}) is given in fig.\,1. The slope
of the curve interpolates between the two values $-8$ and $-12/5$ relative
to the infrared and ultraviolet fixed points; the figure also shows
a very fast ultraviolet convergent pattern, analogously to the massive
cases. The fast convergent behaviour of the series is also confirmed
by the $c$-theorem sum rule
$\Delta c=\frac{3}{2}\int dr\,r^3<\Theta(r)\Theta(0)>
$ \cite{cth}. In fact, we obtain $\Delta c^{(4)}=0.19600\pm .00007$
only including the 4-particle contribution and
$\Delta c^{(6)}=0.1995\pm.0004$ adding the 6-particle one, where the
expected value is $\Delta c=7/10-1/2=0.2$. The fast convergent pattern of
the spectral series also occurs for the 2-point function
of the energy operator $\varepsilon$, with conformal dimensions $(1/10,1/10)$
in the ultraviolet limit and $(1/2,1/2)$ in the infrared one, as shown in
fig.\,2. All its FF can be computed by the recursive equations (\ref{pinch})
with the initial condition $F_{1,1}(\beta)=f_{RL}(\beta)$.

A more subtle situation occurs for the correlation functions of the
magnetization operators. The leading infrared behaviour of the FF of
$\Phi$ and $\tilde\Phi$ can be determined by specializing the FF bootstrap
equations to the Ising critical point where
$S_{RL}=-1$ and $f_{RL}(\be-\be')=e^{(\be-\be')/2}$. With the initial
conditions
$F_{1,0}=F_{0,1}=F_{0,0}=F_{1,1}=1$, the limiting expressions of the FF
are given by
\EQ
F_{r,l}^{IR}(\be_1,\ldots,\be_r;\be_1',\ldots,\be_l')=\prod_{i<j}
\tanh\frac{\be_i-\be_j}{2}\tanh\frac{\be_i'-\be_j'}{2}\,\,,
\lab{ir}
\EN
where $r+l$ is odd for $\Phi$ and even for $\tilde\Phi$.
Expression (\ref{ir}) clearly shows that the
integrals in the spectral representation of
$G_{\Phi}(r)=<\Phi(r)\Phi(0)>$ and $G_{\tilde\Phi}(r)=
<\tilde\Phi(r) \tilde\Phi(0)>$
are infrared divergent, i.e. these correlators require the resummation
of the whole (suitably regularized) spectral series. Although, in general,
it is an interesting open problem to develop mathematical tools that allows
us to deal with such series, here we want to show that an exact resummation
can be performed in the low energy limit so that the exact infrared conformal
dimensions $\Delta_{\Phi}=\Delta_{\tilde\Phi}=1/16$ can be extracted.
In fact, using the FF (\ref{ir}) and exploiting their complete factorization
in a left and a right part we can write
\[
\tilde{G}^{IR}(t)\equiv G^{IR}_{\Phi}(t)+G^{IR}_{\tilde\Phi}(t)=
\left[\sum_{r=1}^\infty\frac{1}{r!}\int^\infty_{-\Lambda}\prod_{i=1}^r
\frac{d\be_i}{2\pi}\left(\prod_{i<j}\tanh\frac{\be_i-\be_j}{2}\right)^2
e^{-t\sum_{i=1}^re^{\be_i}}\right]^2\,\, ,
\]
where $t\equiv Mr/2$ and $\Lambda$ is an infrared cutoff. Shifting
the rapidities, it can be expressed as
\EQ
\tilde{G}^{IR}(t)\,=\,
\left[\sum_{r=1}^\infty\frac{1}{r!}\int^\infty_0\prod_{i=1}^r
\frac{d\be_i}{2\pi}\left(\prod_{i<j}\tanh\frac{\be_i-\be_j}{2}\right)^2
e^{-te^{-\Lambda}\sum_{i=1}^re^{\be_i}}\right]^2\,\,\,.
\EN
As in \cite{CM}, this can be considered as the square of the partition
function of a classical gas of particles living on a semi-infinite line and
subject to the pairwise interaction
$V(\be_i-\be_j)=-\ln\left(\tanh\frac{\be_i-\be_j}{2}\right)^2$.
The activity of such a gas is given by
$z(\be)=\frac{1}{2\pi}e^{-te^{\be-\Lambda}}$. This function has a plateau
$z_0=1/2\pi$ inside a box of length $L\sim\ln\frac{e^\Lambda}{t}$.
Hence, removing the cutoff $\Lambda$ is equivalent to take the thermodynamic
limit $L\goto\infty$ of a gas with a constant activity $z_0$. Standard
relation between the partition function and the bulk-free energy per unit
length $f(z_0)$ allows us to write
$\tilde{G}^{IR}(t)\sim e^{2f\left(\frac{1}{2\pi}\right) L}\sim
\left(\frac{e^\Lambda}{t}\right)^{2f\left(\frac{1}{2\pi}\right)}
$.
The bulk-free energy $f(z_0)$ can exactly be computed for generic values of
$z_0$ \cite{CM},
$f(z_0)=\frac{1}{2\pi}\arcsin(2\pi z_0)-\frac{1}{2\pi^2}\arcsin^2(2\pi z_0)$,
so that $f(1/2\pi)=1/8$, which coincides with the expected infrared anomalous
dimension of the fields $\Phi(x)$ and $\tilde\Phi(x)$.

As explicitly shown in this paper, the FF of massless theories can be exactly
computed and they constitute useful non-perturbative information of the
RG flow. The above analysis also shows that there are sectors
in the theory where the spectral series presents a fast rate of convergence
in the entire domain of definition while for others the power law behaviours
are reproduced by summing the whole series of FF. We expect that this pattern
is also present in other massless QFT and therefore, to take full advantage of
the non-perturbative knowledge of FF, it would be interesting to develop new
resummation methods of the spectral series.

The authors are grateful to F.A. Smirnov and A.B. Zamolodchikov for useful
discussions.

\pagestyle{empty}

\newpage

\hs

\vspace{25mm}

{\bf Figure Captions}

\vspace{1cm}

\begin{description}
\item [Figure 1]. Logarithmic plot of $G_{\Theta}(r)$ including the 4 and
6 particle contributions. Dashed line: leading OPE of the ultraviolet fixed
point.
\item [Figure 2]. Logarithmic plot of the 2-point function of the energy
operator $\varepsilon$ including the 2 and 4 particle contributions.
\end{description}

\end{document}